\documentclass{appolb}
\usepackage{epsfig}

\begin{document}
\title{A sum rule for elastic scattering}
\author{G. Pancheri
\address{Laboratori Nazionali di Frascati, INFN, Frascati, Italy}
\and
Y. Srivastava
\address{Northeastern University, Boston, USA \& INFN, University of Perugia,
Italy}
\and
N. Staffolani
\address{University of Perugia, Perugia, Italy}}

\maketitle
\begin{abstract}
A sum rule is derived for elastic scattering of hadrons at high energies which is in
good agreement with experimental data on $p\bar{p}$ available upto the maximum energy
$\sqrt{s}\ =\ 2 TeV$. Physically, our sum rule reflects the way unitarity correlates
and limits how large the elastic amplitude can be as a function of energy to how fast
it decreases as a function of the momentum transfer. The universality of our result
is justified through our earlier result on equipartition of quark and glue momenta
obtained from the virial theorem for massless quarks and the Wilson conjecture
\end{abstract}
\PACS{PACS numbers come here}

\section{Introduction}

Consider the elastic scattering of two hadrons ($A$ and $B$) with the following
kinematics
$$
A(p_a) + B(p_b)\ \rightarrow\ A(p_c) + B(p_d)
$$
with
$$
s = (p_a + p_b)^2 = (p_c + p_d)^2;
$$
$$
t = (p_a - p_c)^2 = (- p_b + p_d)^2 = - {\vec q}\ ^2 ;
$$
$$
u = (p_a - p_d)^2 = (p_c - p_b )^2, \eqno(1.1)
$$
and let us normalize the elastic amplitude $F(s,t)$ so that the elastic differential
cross-section and the total cross-section (for high energies) read as
$$
({{d\sigma}\over{dt}}) = \pi |F(s,t)|^2;\ \ \sigma_{TOT}(s) = 4\pi \Im m F(s, t=0).
\eqno(1.2)
$$
In the impact parameter representation
$$
F(s,t) = i \int_o^\infty (bdb) J_o(b\sqrt{-t}) {\tilde F}(s,b), \eqno(1.3)
$$
and the partial $b$-wave amplitude is given by
$$
{\tilde F}(s,b) = 1- \eta(s,b)e^{2 i \delta(s,b)}, \eqno(1,4)
$$
where the inelasticity factor $\eta$ lies between ($0 \leq \eta(s,b)\leq 1$) and
$\delta(s,b)$ is the real part of the phase shift. Directly measureable quantities
are (a) $|F(s,t)|$ through $({{d\sigma}\over{dt}})$ and (b) $\Im m F(s,0)$ through
$\sigma_{TOT}$. In the next section 2, we shall obtain lower and upper bounds for a
dimensionless quantity $I_o(s)$ constructed by integrating $|F(s,t)|$ over all
momentum transfers $t$. Under rather mild assumptions, at high energies
($s\rightarrow\ \infty$) it is sharpened into a sum rule
$$
I_o(s) = (1/2) \int_o^\infty(dt)\sqrt{{{d\sigma}\over{\pi dt}}}
= \int_o^\infty(qdq)|F(s,q)|\ \rightarrow\ 1 . \eqno(1.5)
$$
In sec 3, we compare these predictions with the experimental data and find that
already at the Tevatron $\sqrt{s}\ = 2\ TeV$, the integral has the value
$0.98 \pm 0.03$ very close to its asymptotic limit $1$. Our extrapolation for LHC
gives $0.99 \pm 0.03$. Also, a brief discussion of the assumptions and an estimate
of the elastic cross-section is presented here. In sec. IV, we present arguments
based on an equipartition of energy between quark and glue derived earlier, for the
universality of the above result for all hadrons made of light quarks. In the
concluding section, we consider future prospects and possible applications.

\section{Lower and upper bounds and the elastic sum rule}

The dimensionless $b$-wave cross sections are
$$
{{d^2\sigma_{el}}\over{d^2b}} = 1 - 2 \eta(s,b) cos2\delta(s,b) + \eta^2(s,b),
\eqno(2.1a)
$$
$$
{{d^2\sigma_{inel}}\over{d^2b}} = 1 - \eta^2(s,b), \eqno(2.1b)
$$
$$
{{d^2\sigma_{tot}}\over{d^2b}} = 2 [ 1 - \eta(s,b) cos2\delta(s,b) ]. \eqno(2.1c)
$$
The maximum permissible rise for the different cross sections allowed by unitarity
\cite{froissart,martin,cheung,khuri} is when there is total absorption of ``low''
partial waves, i.e., when
$$
\eta(s,b) \rightarrow\ 0,\ as\ b\rightarrow\ 0 \ and\ s\rightarrow\ \infty, \eqno(2.2)
$$
and the ``geometric'' limit is reached
$$
{{d^2\sigma_{el}}\over{d^2b}} = {{d^2\sigma_{inel}}\over{d^2b}} =
(1/2){{d^2\sigma_{tot}}\over{d^2b}}\ \rightarrow\ 1\ (b\rightarrow\ 0;s\rightarrow\
\infty). \eqno(2.3)
$$
Most models with rising total cross-sections satisfy the
above\cite{grau,corsetti,pacetti,godbole,block,gaisser}. Often times, one
defines $\eta(s,b)\ = e^{-n(s,b)/2}$ and $n(s,b)$ is interpreted as the number of
collisions at a given impact parameter $b$ and energy $\sqrt{s}$.

Now let us consider bounds for the dimensionless integral $I_o(s)$ defined in Eq.(1.5).
The lower bound is easily obtained
$$
I_o(s) \geq \int_o^\infty(qdq) |\Im m F(s,q)| \geq \int_o^\infty(qdq) \Im m F(s,q),
\eqno(2.4a)
$$
which upon using Eq.(1.3) leads to
$$
I_o(s) \geq \int(qdq)\int(bdb) J_o(qb) [ 1 - \eta(s,b) cos2\delta(s,b)],
\eqno(2.4b)
$$
so that we have finally
$$
I_o(s) \geq 1 - \eta(s,0) cos2\delta(s,0) \geq 1 - \eta(s,0). \eqno(2.4c)
$$
The upper bound requires more input\cite{staf}. If we assume (an ugly technical
assumption) that $sin2\delta(s,b)$ does not change sign (to leading order in $s$),
then one has the following upper and lower bounds
$$
(1 + {{K}\over{ln(s/s_o)}})\ \geq\ I_o(s)\ \geq\ 1 - \eta(s,0),\ (K\ >\ 0).
\eqno(2.5)
$$
These bounds have been obtained incorporating (i) unitarity, (ii) positivity,
(iii) correct behavior near $b\ =\ 0$ and (iv) the asymptotic behavior for
$b\rightarrow\ \infty$.

Some useful remarks: (1) For hadrons (\it not \rm quarks and glue), the lowest
hadronic state has a finite mass ($m_\pi\ >\ 0$), hence there is a finite range
of interaction. Thus, in the limit of both $b$ and $s$ going to $\infty$, we have
$$
1 - \eta(s,b) cos2\delta(s,b)\ \rightarrow\ 0; \eta(s,b) sin2\delta(s,b)\ \rightarrow
\ 0, \eqno(2.6)
$$
faster than an exponential in $b$. (2) The higher moments
$$
I_n(s) = \int(dt) (-t)^n|F(s,t)|,\ (n\ =\ 1,2,...), \eqno(2.7)
$$
are dimensional and go to zero in the asymptotic limit. Thus, they are less useful
than the zeroeth moment.

From Eq.(2.5), we obtain the sum rule as $s\rightarrow\ \infty$
$$
I_o(s) \rightarrow 1, \ \ as\ s\ \rightarrow\ \infty.\eqno(2.8)
$$

\section{Comparison of the sum rule with experimental data}

 The integral $I_0(s)$ should rise from its threshold value
$2|a_0|k\ \rightarrow\ 0$, where $a_0$ is the S-wave scattering length
(complex for $p\bar{p}$) and $k$ is the CM 3-momentum, to its asymptotic
value $1$ as $s$ goes to infinity.  In Fig.(1), we show a plot of this
integral for available data \cite{abe,amos1800,bozzo,breakstone,amaldi,
akerlof,ayres,asad} on $pp$ and $p\bar{p}$ elastic scattering for
high energies  \cite{bultmann}. Highest energy data at $\sqrt{s}\ =\ 1.8\ TeV$
for $p\bar{p}$ from the Fermilab Tevatron \cite{abe}, give an encouraging
value of $0.98\ \pm\ 0.03$ demonstrating that indeed the integral is close
to its asymptotic value of $1$. We expect it to be even closer to $1$ at
the LHC (our extrapolation gives the value $0.99\pm 0.03$ for LHC).

\vspace{1cm}

\begin{figure}
\begin{center}
\mbox{
\epsfxsize=12truecm \epsffile{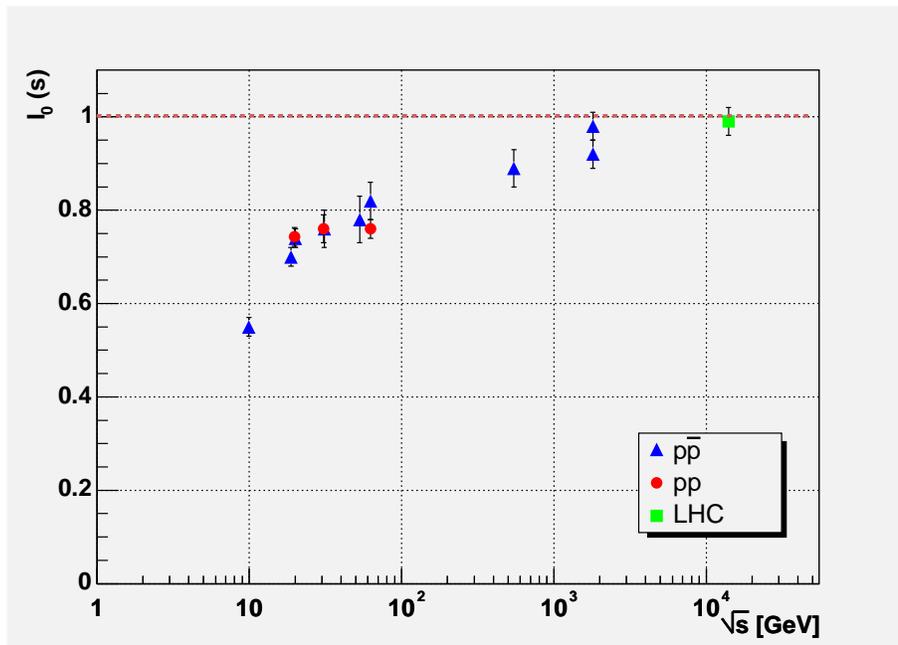}
}
\end{center}
\caption{\label{fig1}A plot of $I_0(s)$ vs. $\sqrt s$ using experimental
data \protect\cite{abe,amos1800,bozzo,breakstone,amaldi,akerlof,ayres,asad}.
The last point is our extrapolation for LHC.}
\end{figure}

\section{ Universality of the sum rule}

It can be shown that the central value of the inelasticity
$\eta(s, 0)\rightarrow\ 0$ at asymptotic energies $s\ \rightarrow\ \infty$
for {\it all} hadrons made of light quarks. Hence, we have the universal
result\cite{staf} that $I_{AB}(s)\ \rightarrow\ 1$ as
$s\ \rightarrow\ \infty$, where $A,B$ are either nucleons or mesons made
of light quarks. The reasons are as follows:
\par\noindent
(i) For nucleons as well as light mesons, half the hadronic energy is
carried by glue. In QCD such an equipartition of energy is rigorously
true\cite{widom, sriv} for hadrons made of massless quarks if the Wilson area
law holds.
\par\noindent
(ii) If we couple (i) to the notion that the rise of the cross-section
is through the gluonic channel, which is flavour independent, the
asynptotic equality of the rise in all hadronic cross-section automatically
emerges.

\section{conclusions}

Our (dimensionless) sum rule reflects the fact that unitarity strongly
correlates the fall off in the momentum transfer to the magnitude of the
scattering amplitude at high energies. Its satisfaction by experimental
data at the highest energy confirms our initial hypotehsis that the
rise in the total cross-section as a function of the energy is indeed
proportional to the fall off in the momentum transfer. As a by product,
we find that the ratio ${{\sigma_{el}}\over{\sigma_{tot}}}\ \rightarrow\
(1/4)$, which is again in very good agreement with data at the highest
Tevatron energy $\sqrt{s}\ =\ 2\ TeV$.

We also find universality. That is, asymptotically, $I_{AB}\ \rightarrow \
1$ for any hadrons $A,B$ made of light quarks. These may be testable at
future LHC and RHIC measurements with heavy ions (or by other means
\cite{bjorken}.

Currently, we are extending similar considerations for one particle inclusive
cross-sections.

\subsection{Acknowledgements}

It is a pleasure to thank Paolo Giromini and Allan Widom for fruitful
discussions. We also take the opportunity to thank Bill Gary and other
organizers of this meeting for their kind hospitality and for making
this conference stimulating and enjoyable.

\end{document}